\documentclass[sigconf, nonacm]{acmart}
\usepackage{subfigure}
\usepackage{color}
\usepackage{algorithm}
\usepackage{algpseudocode}
\usepackage{mathtools}
\usepackage{multirow}
\usepackage{comment}
\usepackage{tikz}
\usetikzlibrary{positioning, shapes.geometric, arrows.meta, calc, decorations.pathreplacing, patterns}
\newcommand{\len}{\text{\texttt{len}}}
\newcommand{\inc}{\text{\texttt{inc}}}

\DeclareMathOperator*{\argmin}{arg\,min}
\usepackage{mathtools, nccmath}

\newcommand{\tol}{\text{\texttt{tol}}}

\definecolor{applegreen}{rgb}{0.55, 0.71, 0.0}

\begin{document}
\title{Parallel Two-Stage Approach for Joint Symbolic Approximation of Time Series}

\author{Xinye Chen}
\orcid{0000-0003-1778-393X}
\affiliation{%
  \institution{Sorbonne Université, CNRS, LIP6}
  \city{Paris}
  \country{France}
}
\email{Xinye.Chen@lip6.fr}

\begin{abstract}
As time-series applications grow larger, there is increasing demand for symbolic representations that are compact, accurate, and scalable across many signals and computing resources. Current ABBA-based symbolic approximation methods produce high-quality, shape-preserving representations, but they handle each time series separately and sequentially. This means they do not ensure consistent symbols across different series and cannot fully exploit modern multicore systems and distributed-memory systems. 
This paper presents a joint symbolic time-series approximation method for large-scale  time series.  The proposed method decouples local compression from global digitization: (i) time series are partitioned into independent domains that can be compressed in parallel, and (ii) the resulting pieces are digitized using a shared global dictionary. To further improve scalability, we introduce a two-stage parallel digitization scheme, in which aggregation is first performed locally and then merged globally without requiring a full-data reassignment step.
Extensive experiments on time-series datasets and large synthetic benchmarks show that our approach maintains competitive reconstruction quality while substantially reducing runtime. These results show that joint symbolic approximation can serve as an efficient, high-level parallel tool for analyzing large-scale temporal data.

\vspace{2pt}


\end{abstract}

\maketitle

\section{Introduction}
Many scientific and industrial applications generate large collections of time series, including sensor monitoring, medical signals, financial data, and high-performance computing traces. These datasets usually have many dimensions, contain noise, and are costly to process in their original form. Because of this, data mining and machine learning algorithms often require substantial computing power when working with raw time-series data, especially as the number of series, channels, or time points increases.  Almost all data mining and machine learning algorithms suffer from the computational load for dealing with large-scale data since they scale poorly with the dimensionality{---}resulting from the curse of dimensionality~\citep{10.1145/276698.276876}. Therefore, it is very desired to compute a representation that reduces the numerosity while preserving the essential characteristics of time series, and the reasonable representation in time series often leads to a boost in algorithmic performance and dramatically alleviates the pressure of compute resources.  

As mentioned in \cite{lin2007experiencing}, representations based on numerical transforms are real-valued, which limits the algorithms and data structures available for them. In this paper, we focus on symbolic approximation techniques. The immediate purpose of symbolic approximation is to perform efficient time series discretization with numerosity reduction, which allows algorithms to manipulate time series in a low-dimensional representation rather than high-dimensional time series features; the efficacy can be verified by, e.g., speeding up neural network inference with symbolic time series representation \citep{EG20b}. However, computing such a representation for large-scale time series is tricky due to its high computational complexity.

The adaptive Brownian bridge-based symbolic aggregation (ABBA) method, as well as its accelerant variant fABBA, is one of the state-of-the-art symbolic approximation techniques regarding reconstruction error in time series domains. However, it requires transforming one time series at a time, which is clumsy for multiple time series, especially at scale. Besides, this method is inherently sequential, which makes it hard to fully utilize available computing resources. More importantly, the consistency of symbols is not guaranteed. The consistency here means each distinct symbol carries the same information in any sample of multiple time series. For example, the symbol ``a'' that appeared in a time series should be identical to the ``a''  in another time series. Besides, the parameter tuning is intractable without prior knowledge, although this problem is already mitigated with fABBA by using tolerance-dominated digitization. Our application of interests focuses on symbolizing multivariate/multiple time series in a unified manner. We propose a joint symbolic representation framework that addresses the aforementioned issues and enables parallelism. The extensive experiments demonstrate that the proposed algorithm can achieve significant speedups while maintaining competitive performance in representation construction, particularly for large-scale time series.  

Our main contributions are summarized as follows:
\begin{enumerate}
    \item We formulate joint symbolic approximation for  time series (including multiple and multivariate time series), where all series share a common dictionary and therefore use consistent symbols.
    
    \item We introduce JABBA, a high-level parallel framework that separates local compression from global digitization. The compression tasks are independent and can be executed concurrently, while the digitization step learns a shared codebook from all compressed pieces.
    
    \item We evaluate the proposed framework on multivariate time-series datasets and synthetic large-scale benchmarks, showing that JABBA achieves substantial runtime reduction while maintaining competitive reconstruction accuracy.
\end{enumerate}

The remainder of this paper is structured as follows. Section~\ref{sec:related} discusses related work of symbolic representation as well as its applications. Section~\ref{sec:pre} reviews the necessary notions of the ABBA framework. Sections~\ref{sec:joint}, ~\ref{sec:pdigit}, and ~\ref{sec:analysis} present our joint method and parallel framework, along with the complexity analysis. Section~\ref{sec:exp} presents the empirical results and analysis, and Section~\ref {sec:final} concludes the paper.

\section{Related work}\label{sec:related}

In this section, we briefly review selected works on symbolic time series representations and their applications. The literature is vast, so we focus on representative methods closely related to our research due to space constraints.

Symbolic representations of time series thus serve as foundational tools in a variety of analysis tasks. Specifically, they play a pivotal role in clustering~\citep{lin2007experiencing, 8933618, 10.1007/s10618-021-00798-w}, classification~\citep{6729617, 8215500, 10.1007/978-3-031-24378-3_4}, forecasting~\citep{EG20b}, event prediction~\citep{7953302}, anomaly detection~\citep{conf/edbt/Senin0WOGBCF15}, and motif discovery~\citep{lin2007experiencing, 10.1145/1814245.1814255, 8970850}.  

Among these symbolic representations, SAX~\citep{lin2003symbolic} was the first to effectively reduce time series dimensionality. It enabled indexing with a lower-bounding distance measure. SAX pioneered a highly influential trend and led to the widespread adoption of symbolic representations in downstream tasks. Such tasks include pattern search (SAXRegEx~\citep{YU202313}), clustering (SAX Navigator~\citep{8933618}; SPF~\citep{10.1007/s10618-021-00798-w}), anomaly detection (HOT SAX~\citep{1565683}; TARZAN~\citep{10.1145/775047.775128}), and classification (SAX-VSM~\citep{6729617}; BOPF~\citep{8215500}; MrSQM~\citep{10.1007/978-3-031-24378-3_4}). Building on SAX, many enhanced variants have emerged, such as 1d-SAX~\citep{10.1007/978-3-642-41398-8_24}, ESAX~\citep{1623910}, pSAX, and cSAX~\citep{9774017}, typically with improvements in speed or accuracy. Yet the original SAX remains widely popular due to its elegant simplicity and efficiency.

ABBA~\citep{EG19b} employs an adaptive polygonal chain approximation followed by mean-based clustering to symbolize time series. Its reconstruction error can be analytically modeled as a Brownian bridge—a random walk with fixed start and end points. The subsequent variant, fABBA~\citep{fABBA2022}, replaces k-means clustering with an efficient greedy aggregation strategy, accelerating the digitization process by orders of magnitude. Both ABBA and fABBA have been empirically shown to better preserve the shape of time series compared to SAX, particularly in capturing local trends such as rises and falls. Applications of ABBA-based representations have proven effective in forecasting and anomaly detection; for instance, LSTM models augmented with ABBA exhibit robust performance across inference tasks~\citep{EG20b}, while TARZAN variants that substitute ABBA or fABBA for SAX outperform the original SAX-based version~\citep{EG19b, fABBA2022}. LLM-ABBA \citep{11397443} demonstrates that the ABBA representation has great potential to enhance downstream time-based applications with large language models. Nevertheless, generating ABBA symbolic representations for multiple time series remains computationally intensive, primarily because of the large number of extracted features and the challenge of ensuring symbolic consistency across series.

Parallel processing is essential for large-scale time-series analytics since modern applications often use many independent signals or high-dimensional streams. A common approach is to decompose the data across time points, variables, or series. These decompositions reveal coarse-grained parallelism and are effective when local computations need minimal synchronization.

However, in symbolic approximation, parallelization presents unique challenges. Compressing each time series independently boosts scalability but yields inconsistent symbols across series. Conversely, learning a global symbolic dictionary ensures consistency but requires aggregating data from all series. To address these trade-offs, the proposed method combines both views: it conducts local compression in parallel, followed by a global digitization step to build a shared dictionary. This design enables efficient multicore and multithreaded execution while preserving a globally consistent symbolic representation.

\section{Preliminaries: ABBA symbolic approximation}\label{sec:pre}

Here we briefly review the ABBA symbolic approximation method. ABBA represents a time series by first computing an adaptive polygonal-chain approximation and then clustering the resulting linear pieces into discrete symbols. The forward symbolization consists of two main steps, compression and digitization, which transform a time series $T = [ t_1, t_2, \ldots, t_n] \in \mathbb{R}^{n}$ into a symbolic approximation 
\begin{equation}\label{eq:symbols}
    A = [ a_1, a_2, \ldots, a_N ],
\end{equation}
where $N \ll n$ and $a_i \in \mathcal{L}$. 

Table~\ref{table:ABBA} summarizes the forward and inverse procedures. In the forward direction, compression transforms the original time series into a sequence of adaptive linear pieces, and digitization maps these pieces to discrete symbols. In the inverse direction, inverse digitization replaces symbols with their representative codebook vectors, quantization restores integer segment lengths, and inverse compression reconstructs the time series. The discrepancy between $T$ and the reconstruction $\widehat{T}$ is referred to as the reconstruction error.

\begin{table}[h]
	\caption{Summarized notation of ABBA procedure} 
	\label{table:ABBA} 
	\centering 
	\small
	\begin{tabular}{l l} 
		\midrule\\[-1mm]
		time series & $T=[t_{0}, t_{1}, \ldots, t_{n}] \in \mathbb{R}^{n}$ \\[1mm]
		after compression & $P=[(\len_{1}, \inc_{1}), \ldots, (\len_{N}, \inc_{N})] \in \mathbb{R}^{2 \times N}$ \\[1mm]
		after digitization & $A=[a_{1}, \ldots, a_{N}] \in \mathcal{L}^{N}$ \\[1mm]
		inverse-digitization & $\widetilde{P}=[(\widetilde{\len}_{1}, \widetilde{\inc}_{1}), \ldots, (\widetilde{\len}_{N}, \widetilde{\inc}_{N})] \in \mathbb{R}^{2 \times N}$ \\[1mm]
		quantization & $\widehat{P}=[(\widehat{\len}_{1}, \widehat{\inc}_{1}), \ldots, (\widehat{\len}_{N}, \widehat{\inc}_{N})] \in \mathbb{R}^{2 \times N}$ \\[1mm]
		inverse-compression & $\widehat{T}=[\widehat{t}_{1}, \widehat{t}_{2}, \ldots, \widehat{t}_{n}] \in \mathbb{R}^{n}$ \\ [1ex]
		\midrule 
	\end{tabular}
\end{table}

\subsection{Compression}
The ABBA compression step aims to compute an adaptive piecewise linear continuous approximation of $T$, that is, to obtain time series pieces $P = [(\len_{1}, \inc_{1}), \ldots, (\len_{N}, \inc_{N})] \in \mathbb{R}^{ N \times 2}$, followed by a reasonable digitization that results in \emph{symbolic sequence} $A = [ a_1, a_2, \ldots, a_N ]\in \mathcal{L}^N$, $N\ll n$, and each $a_j$ is an element of a finite alphabet set $\mathcal{L}$ where $|\mathcal{L}| \ll N$. $\mathcal{L}$ can be referred to as dictionary in the procedure. The ABBA compression adaptively selects $N+1$ indices $i_0 = 0 < i_1 <\cdots < i_N = n$ given a tolerance $\texttt{tol}$ so that the time series $T$ is well approximated by a polygonal chain going through the points $(i_j , t_{i_j})$ for $j=0,1,\ldots,N$. This results in a partition of $T$ into $N$ pieces $p_j=(\len_{j}, \inc_{j})$ that is determined by $T_{i_{j-1}:i_j} = [ t_{i_{j-1}},t_{i_{j-1}+1},\ldots, t_{i_j} ]$, each of integer length $\texttt{len}_j := i_j - i_{j-1}\geq 1$ in the time direction. Visually, each piece $p_j$  is represented by a straight line connecting the endpoint values $t_{i_{j-1}}$ and $t_{i_j}$ This partitioning criterion is the squared Euclidean distance of the values in $p_j$ from the straight polygonal line is upper bounded by $(\texttt{len}_j - 1)\cdot\texttt{tol}^2$. For simplicity, given an index $i_{j-1}$ and starts with $i_0 = 0$, the procedure seeks the largest possible $i_j$ such that $i_{j-1} < i_j\leq n$ and 
\begin{equation}
\begin{aligned}
    \sum_{i=i_{j-1}}^{i_j} \Big( \, t_{i_{j-1}} + (t_{i_j} - t_{i_{j-1}})\cdot \frac{i - i_{j-1}}{i_j - i_{j-1}}  - t_i \Big)^2 \leq (i_{j} - i_{j-1} -1)\cdot\texttt{tol}^2.
    \label{eq:compress}
\end{aligned}
\end{equation}

Each linear piece $p_j$ of the resulting polygonal chain $\widetilde T$ is referred to as a tuple $(\texttt{len}_j, \texttt{inc}_j)$, where $\texttt{inc}_j = t_{i_j} - t_{i_{j-1}}$ is the increment in value, i.e., the subtraction of ending and starting value of $T_{i_{j-1}:i_j}$. The whole polygonal chain can be recovered exactly from the first value $t_0$ and the tuple sequence $p_1, p_2, \ldots, p_N$, i.e.,
\begin{equation}\label{eq:tupseq}
(\texttt{len}_1, \texttt{inc}_1), \ldots, (\texttt{len}_N, \texttt{inc}_N)\in \mathbb{R}^2.
\end{equation}

Because each segment is fixed at its two endpoints, the local approximation error can be interpreted through a Brownian-bridge model.

\subsection{Digitization}
The next step is referred to as digitization, which we further transformed the resulting polygonal chain $\widetilde T$ into the symbolic representation in the form of~\eqref{eq:symbols}.

Following \cite{EG19b}, prior to digitizing, the tuple lengths and increments are separately normalized by their standard deviations $\sigma_{\texttt{len}}$ and $\sigma_{\texttt{inc}}$, respectively. After that, further scaling is employed by using a parameter $\texttt{scl}$ to assign different weights to the length of each piece $p_i$, which denotes importance assigned to its length value in relation to its increment value. Hence, the clustering is effectively performed on the \emph{scaled tuples} 
\begin{equation}
	p_1=\left(\texttt{scl}\frac{\texttt{len}_1}{\sigma_{\texttt{len}}}, \frac{\texttt{inc}_1}{\sigma_{\texttt{inc}}}\right), 
	\ldots, p_B s=\left(\texttt{scl}\frac{\texttt{len}_B}{\sigma_{\texttt{len}}}, \frac{\texttt{inc}_N}{\sigma_{\texttt{inc}}} \right).
	\label{eq:tupseq2}
\end{equation}
In particular, if $\texttt{scl} = 0$, then clustering will be only performed on the increment values of $P$, while if $\texttt{scl} = 1$,  the lengths and increments are clustered with equal importance.

The steps after normalization proceed with a lossy compression technique, e.g., vector quantization (VQ), which is often achieved by mean-based clustering. The concept of vector quantization can be referenced in \cite{1162229, 5075899}. Given an input of $N$ vectors $P=[p_1, \ldots, p_N] \in \mathrm{R}^{\ell\times N}$, VQ seeks a codebook of $k$ vectors, i.e., $C=[c_1, \ldots, c_k] \in \mathrm{R}^{\ell \times k}$ such that $k$ is much smaller than $N$ where each $c_i$ is associated with a unique cluster $S_i$. A quality codebook enables the sum of squared errors $\texttt{SSE}$ to be small enough to an optimal level. Suppose $k$ clusters $S_1, S_2, \ldots, S_k \subseteq P$ are computed, VQ aims to minimize
\begin{equation}\label{eq:sse}
    \texttt{SSE} = \sum_{i=1}^{k} \phi (c_i, S_i) = \sum_{i=1}^{k}\sum_{p\in S_i}\mathrm{dist}(p, c_i)^2,
\end{equation}
where $\phi$ denotes energy function, $c_i$ denotes the center of cluster $S_i$ and $\mathrm{dist}(p_i, p_j)$ often denotes the Euclidean norm $\|p_i - p_j\|_2$. We often choose the mean center $\mu_i$ as $c_i$ for Euclidean space, i.e., $\mu_i = \frac{1}{|S_i|}\sum_{p \in S_i} p$, and then \eqref{eq:sse} can be written as $\texttt{SSE} = \sum_{i=1}^{k} |S_i| \text{Var} S_i$.  Lloyd's algorithm \citep{journals/tit/Lloyd82} (also known as k-means algorithm) is a suboptimal solution of vector quantization to minimize $\texttt{SSE}$.

The ABBA digitization can be performed by a suitable partitional clustering algorithm that finds $k$ clusters from $P \in \mathbb{R}^{N \times 2}$ such that the sum of Euclidean distance $\texttt{SSE}$ constructed by $C$ is minimized. The obtained codebook vectors are referred to \emph{symbolic centers} here. Each symbolic center is associated with an identical symbol and each time series snippet $p_i$ is assigned with the closest symbolic center $c^i$ associated with its symbol 
\begin{equation}
    c_i^\star = \argmin_{c \in C} \|p_i-c\|_2 .
\end{equation}

The symbolic centers to symbols are one-to-one mapping, denoted by $I_d: C \to A$ , thus the digitization $f_d: P \to A$ is given by  
\begin{equation}
    f_d(p_i)=I_d(c_i^\star).
\end{equation}

The greedy aggregation is introduced in \cite{fABBA2022}, which first sorts the data and then greedily aggregates it into groups. The sorting order naturally avoids unnecessary computations in aggregation by triggering an early stopping. The codebook set is constructed from the aggregated group means, serving as a suboptimal solution to the k-means problem. Though its accuracy is less significant than Lylod's algorithm, it achieves a significant speedup, and the $\texttt{SSE}$ is upper-bounded by $\alpha^2 (N - k)$ for $N$ data points.  We assume variance of length and increment of pieces, denoted by $\texttt{Var}_{\texttt{len}}$ and $\texttt{Var}_{\texttt{inc}}$, are:
\begin{equation}\label{eq:Var}
    \begin{aligned}
        &\texttt{Var}_{\texttt{len}} = \max_{i = 1,\ldots,k}\frac{1}{|S_{i}|}\sum_{\texttt{len}\in S_{i}}|\texttt{len} - \mu_i^{\texttt{len}}|^2, \\
        &\texttt{Var}_{\texttt{inc}} = \max_{i = 1,\ldots,k}\frac{1}{|S_{i}|}\sum_{\texttt{inc}\in S_{i}}|\texttt{inc} - \mu_i^{\texttt{inc}}|^2.
    \end{aligned}
\end{equation}

Here we suppose the aggregation is performed on the length and increment values (1-dimensional data) of pieces simultaneously, which is referred to as \emph{hierarchical aggregation}, and we denote the digitization tolerance for length and increment $\alpha_{\texttt{len}}$ and  $\alpha_{\texttt{inc}}$, respectively. Obviously, we have 
\begin{equation}\label{eq:Var1}
    \max(\texttt{Var}_{\texttt{len}}, \texttt{Var}_{\texttt{inc}}) \le \max(\alpha_{\texttt{len}}, \alpha_{\texttt{inc}})^2
\end{equation}

Using aggregation, we have $\alpha_{\texttt{len}} = \alpha_{\texttt{inc}} =  \alpha$, this yields 
\begin{equation}\label{eq:Var2}
    \max(\texttt{Var}_{\texttt{len}}, \texttt{Var}_{\texttt{inc}}) \le \alpha^2
\end{equation}

Each symbol is associated with a unique cluster. In practice, each clustering label (membership) corresponds to a unique byte-size integer value.  The symbols used in ABBA can be represented by text characters, which are not limited to English alphabet letters {---}; often, more clusters will be used. Each character in most computer systems is represented by an ASCII string with a unique byte-size integer value (a unique cluster membership). Besides, it can be any combination of symbols or an ASCII representation.  The digitization is the key to compression rate, which is the size of codebook $C$ (i.e., the number of distinct symbols $|\mathcal{L}|$) divided by the length of the time series. 
 

\subsection{Inverse symbolization}\label{subsec:inverse_sb}
The inverse symbolization refers to the process from $A$ to $\widehat{T}$, the intuition is to reconstruct time series from \eqref{eq:symbols} such that the reconstructed time series $\widehat{T}$ is as close to $T$ as possible.  The inverse symbolization contains three steps. 

The first step is referred to as \emph{inverse-digitization}, simply written as $f_d^{-1}$, which uses the $k$ representative elements $c_i$ (in terms of, e.g., mean centers or median center of the groups $S$) from codebook $C$ to replace the symbol in $A$ orderly, and thus results in a 2-by-$N$ array $\widetilde{P}$, i.e., an approximation of $P$, where each $\widetilde{p}_i \in \widetilde{P}$ is the closest symbolic center $c \in C$ to $p_i \in P$. The inverse digitization often leads to a non-integer value to the reconstructed length $\len$, so \cite{EG19b} proposes a novel rounding method, which is referred to as  \emph{quantization}, to align the cumulated lengths with the closest integers. The method is as follows: start with rounding the first length into an integer value, i.e., $\widehat{\len}_1:= \text{round}(\widetilde{\len}_1)$ and calculate the rounding error $e:= \widetilde{len}_1 - \widehat{len}_1$. The the error is added to the rounding to $\widetilde{\len}_2$, i.e., $\widehat{\len}_2 := \text{round}(\widetilde{\len}_2 + e)$ and new error $e'$ is calculated as $\widehat{\len}_2 + e - \widetilde{\len}_2$. Then $e'$ is involved in the next rounding similarly. After all rounding is computed, we obtain 
\begin{equation}\label{eq:repolygon}
\widehat{P}=[(\widehat{\len}_{1}, \widehat{\inc}_{1}), \ldots, (\widehat{\len}_{N}, \widehat{\inc}_{N})] \in \mathbb{R}^{2 \times N},
\end{equation}
where increments $\inc$ are unchanged, i.e., $\widehat{\inc} = \widetilde{\inc}$. Then, the whole polygonal chain can be recovered exactly from the initial time value $t_0$ and the tuple sequence \eqref{eq:repolygon} via the inverse-compression.

The lower reconstruction error means a higher approximation accuracy. The reconstruction error can be defined by mean squared error (MSE), which is given by 
\begin{equation}
    \text{MSE}=\frac{1}{n}\sum_{i=1}^{n}(t_i-\widehat{t}_i)^2.
\end{equation}

\section{Joint symbolic approximation}\label{sec:joint}
The design of JABBA follows a simple observation: ABBA compression is sequential within one time series, but independent across different time series or disjoint partitions. Therefore, large-scale symbolization can expose parallelism by decomposing the input into independent domains. The challenge is that independent compression and digitization would produce inconsistent symbols. JABBA resolves this by performing compression locally but digitization globally. This preserves a shared symbolic dictionary while allowing the computationally expensive compression stage to be parallelized.


The ideal case of symbolization of multiple time series is that the symbolization should have consistent symbols used in each time series and as less distinct symbols used as possible. One intuitive idea is to fit one (or a given number of) time series and use the previous symbolic information to transform the rest of the data. However, it does not consider the variety of characteristics in every single time series, and this might result in serious information loss in some time series. Henceforth, we require an approach, i.e., joint symbolic approximation, that can symbolize multiple time series simultaneously.

The essential idea of JABBA is to decouple the compression and digitization stages while preserving a consistent symbolic representation across all time series. Let $\mathcal{T}$ be a dataset of $M$ time series. When $M=1$, the same framework applies by partitioning the series into multiple subsequences.

A naive parallelization would perform both compression and digitization independently on each partition, leading to inconsistent symbol assignments across different series. To address this, we adopt a two-stage strategy: compression is performed independently on each series (or partition), producing local piece sequences, while digitization is carried out globally on the union of all pieces. This ensures that all time series share a common symbolic dictionary. 

This design enables parallelism in the computationally expensive compression stage while maintaining global consistency in the symbolic representation. The joint symbolic approximation framework, together with the parallel computing paradigm, is illustrated in \figurename~\ref{fig:PA}, and described in Algorithm~\ref{algo:j_abba}.



Asa result, the approach can be applied to datasets storing multiple time series such as UCR time series archive \citep{UCRArchive2018}. With the availability of consistent symbols information, techniques of text mining and natural language processing are becoming promising in time series analysis. 

 \begin{figure*}[ht]
	\centering
\begin{tikzpicture}[
    font=\sffamily,
    >=Stealth,
    main box/.style={
        draw=black!70, 
        thick, 
        fill=white, 
        rounded corners=2pt, 
        minimum height=1.2cm, 
        align=center
    },
    process box/.style={
        draw=blue!80!black, 
        thick, 
        fill=blue!5, 
        rounded corners=3pt, 
        minimum height=1.5cm, 
        minimum width=2.5cm, 
        align=center,
        font=\small
    },
    data node/.style={
        draw=none, 
        align=center, 
        font=\small\bfseries
    },
    arrow/.style={
        ->, 
        thick, 
        color=black!70
    }
]

    \node[data node] (title) at (0, 0.5) {Original Time Series $T$};
    
    \draw[thick, ->] (-6, 0) -- (6, 0) node[right] {$t$};
    \draw[blue!70!black, thick, smooth] plot coordinates {
        (-5.5, 0.2) (-5, 0.8) (-4, -0.3) (-3, 0.5) 
        (-2, 0.2) (-1, -0.6) (0, 0.4) (1, 0.8) 
        (2, -0.2) (3, 0.6) (4, 0.1) (5, 0.9) (5.5, 0.2)
    };

    \draw[red, dashed, thick] (-2, -0.5) -- (-2, 1);
    \draw[red, dashed, thick] (2, -0.5) -- (2, 1);

    \draw[decorate, decoration={brace, amplitude=5pt, mirror}, thick, gray] 
        (-5.8, -0.6) -- (-2.1, -0.6) node[midway, below=5pt] {$T^{(1)}$ (Chunk 1)};
        
    \draw[decorate, decoration={brace, amplitude=5pt, mirror}, thick, gray] 
        (-1.9, -0.6) -- (1.9, -0.6) node[midway, below=5pt] {$T^{(m)}$ (Chunk $m$)};
        
    \draw[decorate, decoration={brace, amplitude=5pt, mirror}, thick, gray] 
        (2.1, -0.6) -- (5.8, -0.6) node[midway, below=5pt] {$T^{(M)}$ (Chunk $M$)};

    \node[process box] (proc1) at (-4, -3) {
        \textbf{Core 1} \\ 
        Local Compress \\ 
        $(P^{(1)}) = \Psi(T^{(1)})$
    };
    
    \node[process box] (prock) at (0, -3) {
        \textbf{Core $m$} \\ 
        Local Compress \\ 
        $(P^{(m)}) = \Psi(T^{(m)})$
    };
    
    \node[process box] (procM) at (4, -3) {
        \textbf{Core $M$} \\ 
        Local Compress \\ 
        $(P^{(M)}) = \Psi(T^{(M)})$
    };

    \draw[arrow] (-4, -1.2) -- (proc1);
    \draw[arrow] (0, -1.2) -- (prock);
    \draw[arrow] (4, -1.2) -- (procM);
    
    \node at (-2, -3) {\Large $\dots$};
    \node at (2, -3) {\Large $\dots$};

    \node[main box, minimum width=10cm, fill=green!5, draw=green!60!black] (result) at (0, -5.5) {
        \textbf{Concatenation (Reduce)} \\
        $P = P^{(1)} \oplus \dots \oplus P^{(m)} \oplus \dots \oplus P^{(M)}$
    };

    \draw[arrow] (proc1.south) -- (result.north -| proc1.south);
    \draw[arrow] (prock.south) -- (result.north -| prock.south);
    \draw[arrow] (procM.south) -- (result.north -| procM.south);

    \node[right=0.2cm of title, text width=4cm, font=\scriptsize, gray] (note) {
        \textcolor{red}{-- --} Partition Boundaries \\
        (Potential Artifacts)
    };

\end{tikzpicture}
\caption{The joint ABBA framework.}\label{fig:PA}
\end{figure*}

\begin{algorithm}[H]
\caption{JABBA: Parallel joint symbolic approximation}
\label{algo:j_abba}
\small
\begin{algorithmic}[1]
\Require Input time-series collection $\mathcal{T}$, compression tolerance $\texttt{tol}$, digitization method $\mathcal{C}$, number of partitions/workers $M$
\Ensure Symbolic representations $\mathcal{A}=\{A^{(1)},\ldots,A^{(M)}\}$ and shared dictionary $\mathcal{D}$

\Statex
\Statex \textbf{\textsc{Phase I: Domain construction}}
\State Construct domains $\{T^{(1)},\ldots,T^{(M)}\}$ from $\mathcal{T}$
\Comment{series, channels, or chunks}

\Statex
\Statex \textbf{\textsc{Phase II: Parallel compression}}
\For{$k=1,\ldots,M$ \textbf{in parallel}}
    \State $P^{(k)} \gets \textsc{Compress}(T^{(k)}, \texttt{tol})$
\EndFor

\Statex
\Statex \textbf{\textsc{Phase III: Global digitization}}
\State $P_{\mathrm{global}} \gets P^{(1)} \cup \cdots \cup P^{(M)}$
\State $\mathcal{D} \gets \mathcal{C}(P_{\mathrm{global}})$

\For{$m=1,\ldots,M$}
    \State $A^{(m)} \gets \textsc{Digitize}(P^{(m)}, \mathcal{D})$
\EndFor

\Statex
\State \Return $\mathcal{A}, \mathcal{D}$
\end{algorithmic}
\end{algorithm}

\subsection{Parallelization via Domain Decomposition}

To mitigate the sequential dependency inherent in the greedy strategy without incurring the computational overhead of exhaustive reachability mapping, we propose a \emph{Domain Decomposition} approach. This method relaxes the global optimality constraint at specific boundaries to expose coarse-grained parallelism while keeping synchronization minimal.

\subsubsection{Data Partitioning}
The input time series $T$ of length $n$ is partitioned into $M$ disjoint sub-sequences (or \emph{chunks}) $\{T^{(1)}, T^{(2)}, \ldots, T^{(M)}\}$. Assuming a balanced load distribution, the domain indices are split uniformly, such that the $k$-th chunk $T^{(k)}$ covers the index range $[s_k, e_k]$, where $s_1=0$, $e_M=n$, and $s_{k+1} = e_k$.
\begin{equation}
    T^{(k)} = [t_{s_k}, t_{s_k+1}, \ldots, t_{e_k}], \quad \text{for } k=1, \ldots, M.
\end{equation}

\subsubsection{Local Parallel Compression}
The compression operator $\Psi$, defined by the greedy ABBA procedure (Eq.~\eqref{eq:compress}), is applied independently to each sub-sequence. Let $P^{(m)}$ and $A^{(m)}$ denote the pieces and symbolic sequence resulting from the compression of the $m$-th chunk:
\begin{equation}
    (P^{(m)}, A^{(m)}) = \Psi(T^{(m)}, \texttt{tol}), \quad \text{computed in parallel}.
\end{equation}
Since the computation of $\Psi(T^{(m)})$ depends solely on the data within $[s_k, e_k]$, these tasks are strictly independent and can be executed simultaneously on $M$ processing units without communication during the local compression phase.

\subsubsection{Sequence concatenation and boundary effects}
The global approximation is constructed by the concatenation of the local results. The final tuple sequence $P$ is given by:
\begin{equation}
    P = P^{(1)} \oplus P^{(2)} \oplus \cdots \oplus P^{(M)},
\end{equation}
where $\oplus$ denotes the concatenation operator. It is important to note that this approach introduces \emph{boundary artifacts}. Specifically, a linear trend in the original series $T$ that spans across a partition boundary $e_k$ will be forcibly split into two separate segments in $P$, one ending at $e_k$ and the other starting at $s_{k+1}$. When a long univariate series is partitioned, a segment that would have crossed a partition boundary in the serial algorithm must be split into local segments. This may slightly increase the number of pieces and change the reconstruction near the boundaries. The number of boundary-induced pieces grows with the number of partitions. In large datasets, where the number of pieces is much larger than the number of partitions, this effect is typically small compared with the scalability benefit.

\section{Two-stage Parallel digitization}\label{sec:pdigit}
The runtime of ABBA symbolization is dominated by the compression phase, where we have an efficient parallel algorithm. In this section, we discuss the potential speedup brought by the parallel aggregation.  

The sequential nature of greedy aggregation arises from the fact that each starting point determines the assignment of subsequent points. However, a key observation is that the algorithm operates on data sorted by a scalar value (e.g., norm or PCA projection), and uses the pruning condition
\begin{equation}
    \mathrm{sort}(p_j) - \mathrm{sort}(p_i) > \alpha,
\end{equation}
to terminate neighborhood search. This implies that points whose sorting values differ by more than the tolerance $\alpha$ cannot belong to the same group. This locality property allows the aggregation process to be approximated in parallel by partitioning the sorted data into disjoint blocks.

We design a two-stage parallel aggregation algorithm implemented, which decomposes the digitization process into local aggregation and global merging.  

\paragraph{Stage I: Parallel local aggregation.}
Let $P = \{p_1, \ldots, p_N\}$ denote the set of pieces obtained after compression. We first compute a global sorting index and partition the sorted data into $M$ contiguous blocks
\[
P = P^{(1)} \cup P^{(2)} \cup \cdots \cup P^{(M)}.
\]
Each MPI worker independently applies the aggregation procedure to its local block:
\[
(\ell^{(m)}, S^{(m)}) = \textsc{Aggregate}_{\mathrm{presorted}}(P^{(m)}, \{\rho_i\}_{i\in \pi^{(m)}}, \alpha),
\]
where $\ell^{(m)}$ are the local labels and $S^{(m)}$ is the set of local starting points.

Importantly, each block $P^{(m)}$ preserves the global sorted order, and the aggregation is performed without re-sorting. The pruning condition based on sorting values is therefore still valid locally.

Here $\textsc{Aggregate}_{\mathrm{presorted}}$ denotes the aggregation procedure applied to data that is already sorted according to the global ordering, without re-sorting, while still using the pruning condition based on $\rho$. Since each worker only accesses its local data, this stage is embarrassingly parallel and scales linearly with the number of processes.

\paragraph{Stage II: Global merging of starting points.}
After local aggregation, all starting points are collected:
\[
S_{\mathrm{global}} = \bigcup_{m=1}^M S^{(m)}.
\]
We then apply the same aggregation procedure to $S_{\mathrm{global}}$:
\[
\tilde{\ell} = \textsc{Aggregate}(S_{\mathrm{global}}, \alpha),
\]
which produces a mapping from local groups to global groups.

\paragraph{Label propagation.}
Instead of recomputing distances between all data points and the merged centers, we directly update labels via group mapping. For a point $p_i$ belonging to a local group $g$, its final label is given by
\begin{equation}
    \ell_i = \tilde{\ell}_{g}.
\end{equation}
This avoids a second pass over the entire dataset for distance computation.

\section{Complexity and scalability}\label{sec:analysis}

Let $n$ denote the length of the original time series and $N$ the number of pieces after compression. Let $G$ denote the number of starting points generated during digitization, where typically $G \ll N$. We use $C_{\mathrm{comp}}(n)$ to denote the cost of compression and $C_{\mathrm{agg}}(N)$ to denote the cost of greedy aggregation on $N$ pieces.

\paragraph{Serial complexity.}
In the standard ABBA pipeline, the total computational cost is
\begin{equation}
    C_{\mathrm{serial}} = C_{\mathrm{comp}}(n) + C_{\mathrm{agg}}(N),
\end{equation}
where the digitization step is implemented via aggregation.

\paragraph{Parallel compression.}
With $M$ balanced partitions and $M$ workers, the compression stage can be parallelized as
\begin{equation}
    C_{\mathrm{comp}}^{\mathrm{parallel}} \approx \max_{m=1,\ldots,M} C_{\mathrm{comp}}(n_m),
\end{equation}
where $n_m \approx n/M$ is the size of the $m$-th partition.

\paragraph{Parallel digitization.}
For the proposed two-stage aggregation, the digitization cost becomes
\begin{equation}
    C_{\mathrm{agg}}^{\mathrm{parallel}} \approx \max_{m=1,\ldots,M} C_{\mathrm{agg}}(|P^{(m)}|) + C_{\mathrm{agg}}(G),
\end{equation}
where $|P^{(m)}| \approx N/M$ is the number of pieces in each partition, and each local aggregation operates on presorted data without additional sorting cost.

Unlike a naive parallelization that requires a full reassignment step with cost $O(NM)$, the proposed method performs distance computations only in: (1) local aggregation within each partition, and (2) aggregation over the starting points. Therefore, the number of distance evaluations is reduced from $O(NM)$ to approximately $O(N + G)$.

Combining both stages, the total parallel cost is
\begin{equation}
    C_{\mathrm{parallel}} \approx \max_{m} C_{\mathrm{comp}}(n_m) + \max_{m} C_{\mathrm{agg}}(|P^{(m)}|) + C_{\mathrm{agg}}(G).
\end{equation}
The achievable speedup is therefore governed by the fraction of computation spent in the parallelizable components, in accordance with Amdahl's law. In particular, the removal of a global reassignment step and the preservation of sorting-based pruning within each block further improve practical scalability.

\section{Empirical Results}\label{sec:exp}

In this section, we focus on experiments on runtime and reconstruction errors in symbolic representation.  We conduct extensive experiments on the synthetic Gaussian blobs, the UEA Archive \citep{UEAArchive2018}, and Gaussian noises for the multithreading test. We select the competing algorithms that provide publicly available software\footnote{Available at \url{https://github.com/nla-group/ABBA} and \url{https://github.com/nla-group/fABBA}.}, which is for simplicity and efficiency. 

\subsection{Simulations on parallel aggregation}

We test the MPI-based two-stage aggregation scheme to speed up digitization. The experiment looks at how well the method scales and how the approximation affects clustering results on Gaussian blob data.

We create synthetic two-dimensional datasets with sizes $N \in \{5{,}000, 10{,}000\}$. The tolerances $\alpha$ are set to $0.1$ and $0.05$ to adjust the level of aggregation detail. We compare serial aggregation (Serial GA) with the MPI two-stage method using $M=4$ and $M=8$ workers. In the parallel setup, data are sorted and split, aggregated locally, and then merged in a second aggregation step.

Table~\ref{tab:mpi_aggregate_combined} shows the results. As expected, using more workers lowers the runtime when the dataset is large enough. For instance, with $\alpha=0.1$ and $N=5{,}000$, $M=8$ gives a speedup of over $15\times$. In smaller setups, the improvement is less consistent due to communication overhead. The MPI method slightly increases SSE and reduces the number of clusters, since block-wise processing can merge points that would stay separate in the serial version. This effect is smaller when $\alpha$ is lower, making clustering finer and more stable.

The MPI-based parallelism for aggregation clearly reduces runtime while preserving most of the clustering structure, demonstrating that the approximation works well in practice.

\begin{table*}[h]
\centering
\caption{MPI two-stage aggregation scaling under different $\alpha$ and data sizes.}
\label{tab:mpi_aggregate_combined}\setlength\tabcolsep{2.pt}
\begin{tabular}{c c c c c c c c c}
\toprule
$\alpha$ & $N$ & Method & $M$ & Time (s) & Speedup & SSE ratio & \#Clusters (Serial) & \#Clusters (MPI)\\
\midrule

0.1 & 5000 
    & Serial GA      & -- & 0.0035 & 1.000 & 1.0000 & 1,153 & -- \\
    & 
    & MPI two-stage  & 4  & 0.0213 & 1.053 & 1.8326 & 1,153 & 1,051 \\
    & 
    & MPI two-stage  & 8  & 0.0046 & 15.714 & 1.7239 & 1,153 & 1,096 \\

\midrule

0.1 & 10000 
    & Serial GA      & -- & 0.0038 & 1.000 & 1.0000 & 1,492 & -- \\
    & 
    & MPI two-stage  & 4  & 0.0281 & 1.356 & 2.1394 & 1,492 & 1,342 \\
    & 
    & MPI two-stage  & 8  & 0.0107 & 8.933 & 2.0128 & 1,492 & 1,366 \\

\midrule

0.05 & 5000 
     & Serial GA      & -- & 0.0062 & 1.000 & 1.0000 & 2,371 & -- \\
     & 
     & MPI two-stage  & 4  & 0.0202 & 1.815 & 1.3978 & 2,371 & 2,293 \\
     & 
     & MPI two-stage  & 8  & 0.0092 & 6.666 & 1.2690 & 2,371 & 2,322 \\

\midrule

0.05 & 10000 
     & Serial GA      & -- & 0.0099 & 1.000 & 1.0000 & 3,419 & -- \\
     & 
     & MPI two-stage  & 4  & 0.0136 & 3.032 & 1.6146 & 3,419 & 3,226 \\
     & 
     & MPI two-stage  & 8  & 0.0185 & 3.610 & 1.4500 & 3,419 & 3,270 \\

\bottomrule
\end{tabular}
\end{table*}

\subsection{Simulations on multivariate time series}
The UEA Archive contains 30 multivariate time series datasets with a variety of dimensions and lengths. The datasets are very huge, therefore it is inefficient for the original ABBA and its variant fABBA to perform computations one at a time, and the symbols are not consistent across channels. 

To verify the effectiveness of the digitization following the last experiment, we evaluate the proposed MPI-based digitization using $K=4$ and $K=8$ workers, which correspond to typical multi-core configurations on modern CPUs.

\begin{table*}[h]
\caption{MPI scaling of JABBA digitization on selected UEA multivariate time-series datasets. Runtime denotes digitization time.}
\label{tab:UEA_mpi_scaling}
\centering
\setlength\tabcolsep{16pt}
\begin{tabular}{c c c c}
\toprule
Dataset & Metric & $M=4$ & $M=8$ \\
\midrule
\multirow{5}{*}{\shortstack{BasicMotions\\Size=80, Dim=6, Len=100\\($\tol=0.01$)}} 
& MSE & \textbf{2.18} & 2.29 \\
& Digit. time & \textbf{0.037} & 0.038 \\
& Speedup & 1.00x & \textbf{1.45x} \\
& Efficiency & \textbf{0.25} & 0.18 \\
& Symbols & 210 & 215 \\
\midrule

\multirow{5}{*}{\shortstack{CharacterTrajectories\\Size=2858, Dim=3, Len=182\\($\tol=0.01$)}} 
& MSE & 0.706 & \textbf{0.628} \\
& Digit. time & \textbf{0.177} & 0.180 \\
& Speedup & 0.89x & \textbf{1.16x} \\
& Efficiency & \textbf{0.22} & 0.15 \\
& Symbols & 328 & 339 \\
\midrule

\multirow{5}{*}{\shortstack{Epilepsy\\Size=275, Dim=3, Len=206\\($\tol=0.11$)}} 
& MSE & \textbf{7.69} & 7.7 \\
& Digit. time & \textbf{0.095} & 0.133 \\
& Speedup & \textbf{1.01x} & 0.93x \\
& Efficiency & \textbf{0.25} & 0.12 \\
& Symbols & 487 & 497 \\
\midrule

\multirow{5}{*}{\shortstack{NATOPS\\Size=360, Dim=24, Len=51\\($\tol=0.01$)}} 
& MSE & \textbf{0.0828} & 0.104 \\
& Digit. time & \textbf{0.193} & 0.250 \\
& Speedup & 0.99x & \textbf{1.05x} \\
& Efficiency & \textbf{0.25} & 0.13 \\
& Symbols & 535 & 559 \\
\midrule

\multirow{5}{*}{\shortstack{StandWalkJump\\Size=27, Dim=4, Len=2500\\($\tol=0.01$)}} 
& MSE & \textbf{24.47} & 40.59 \\
& Digit. time & \textbf{0.038} & 0.046 \\
& Speedup & \textbf{1.01x} & 0.96x \\
& Efficiency & \textbf{0.25} & 0.12 \\
& Symbols & 455 & 471 \\
\midrule

\multirow{5}{*}{\shortstack{UWaveGestureLibrary\\Size=440, Dim=3, Len=315\\($\tol=0.01$)}} 
& MSE & \textbf{0.257} & 0.316 \\
& Digit. time & \textbf{0.026} & 0.033 \\
& Speedup & 1.29x & \textbf{2.66x} \\
& Efficiency & 0.32 & \textbf{0.33} \\
& Symbols & 583 & 609 \\

\bottomrule
\end{tabular}
\end{table*}

To isolate the performance of the digitization stage, we maximize parallelism in the compression phase by assigning the maximum number of available jobs. This ensures compression does not become the computational bottleneck, allowing a fair assessment of the scalability of the digitization procedure.

Table~\ref{tab:UEA_mpi_scaling} reports the performance of the proposed MPI-based digitization for $M=4$ and $M=8$ workers. Overall, the observed speedup is moderate, and in some cases close to $1\times$, indicating limited scalability of this stage in isolation.

There are two main reasons for this result. First, the digitization step works on compressed pieces, and their size $N$ is usually much smaller than the original time series. This means each worker has less work to do, so communication and synchronization overhead limit parallel efficiency. Second, the algorithm needs a global merging step for the starting points, which adds sequential parts and further limits scalability as described by Amdahl’s law.

We also observe that using more workers, from $M=4$ to $M=8$, does not always make the process faster and can even slow it down for smaller datasets. This happens because communication costs rise and each worker has less to do, thereby lowering parallel efficiency.  Even though speedup is limited, the reconstruction quality (MSE) and the number of symbols stay about the same for different values of $M$. This shows that parallelization does not have much effect on digitization quality.

It is important to note that digitization is not the main computational bottleneck in the ABBA pipeline. Most of the benefit from parallelization comes from the compression stage, since its workload grows with the time series length. The MPI-based digitization is best seen as a supporting part that helps achieve full parallelism, not as a main source of speedup. 

\begin{figure*}
    \centering
    \includegraphics[width=0.7\linewidth]{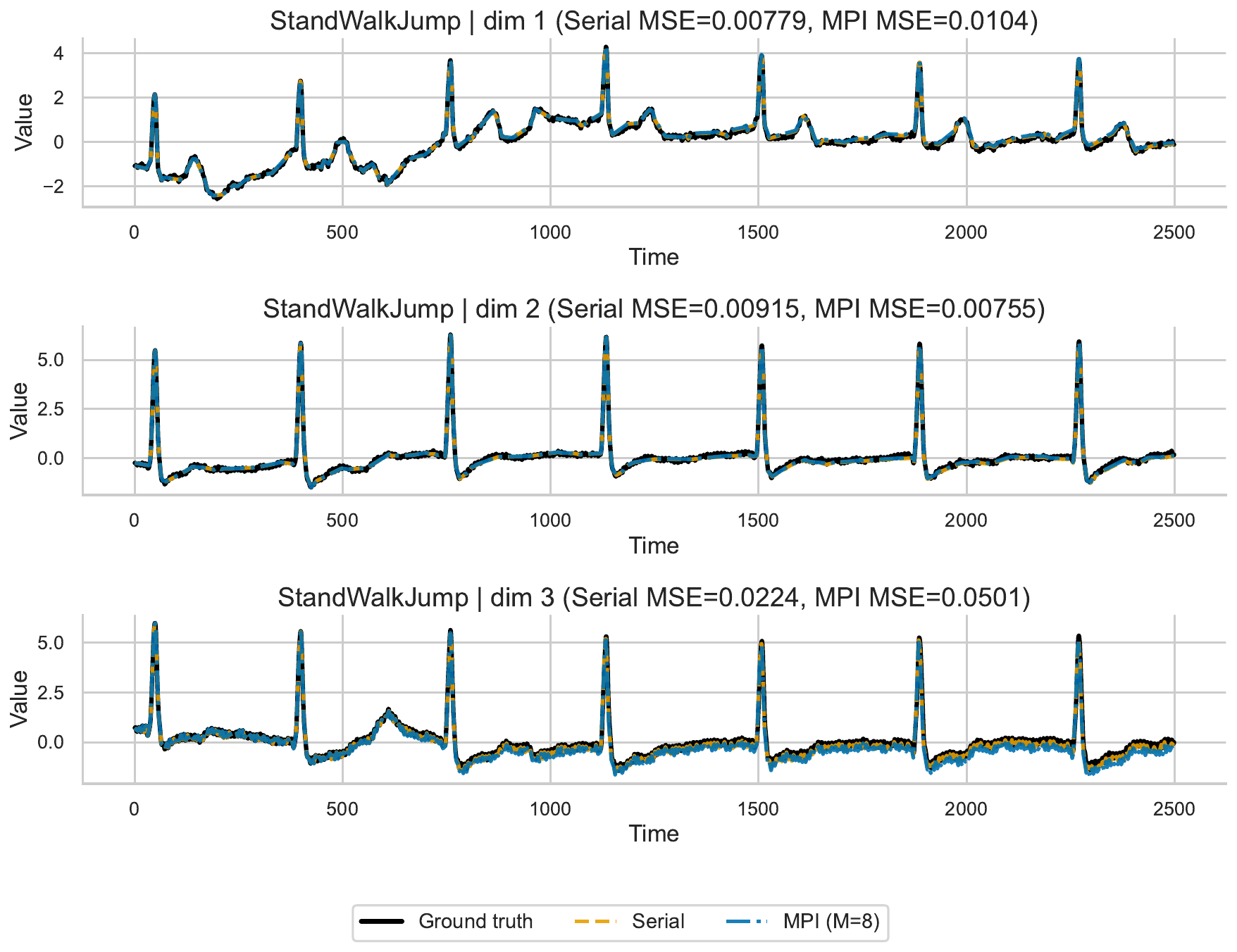}
    \caption{Comparison of reconstruction; We visualize the first three dimensions of the first sample from StandWalkJump dataset. }
    \label{fig:StandWalkJump}
\end{figure*}

Across the tests, the JABBA achieves comparable MSE to the serial version with a similar number of symbols and high-quality reconstruction. \figurename~\ref{fig:StandWalkJump} shows the reconstruction between ground truth time series and JABBA with serial computing and MPI parallelism ($M=8$), showing the competitiveness of JABBA's accuracy, and that other datasets obtain similar results.

\subsection{Multithreading simulation on univarate time series}
In this experiment, we will compare ABBA, fABBA, JABBA on synthetic Gaussian noise series in terms of runtime, and reconstruction accuracy with various number of time series partitions. The reconstruction accuracy is measured by MSE here. We only parallelize the compression stage, while keeping the digitization stage sequential. 

We used Gaussian noises as the time series for benchmarking. The data generated for the test are of length 100,000 with zero mean and unit standard deviation.  We first ran fABBA with $\tol=0.01$ and $\alpha=0.05$ to compute the number of symbols it used. This simulation used 358 symbols accordingly. Second, we ran ABBA by feeding the same number of symbols fABBA used to $k$, i.e., $358$ symbols. After that, we run the JABBA  with varying partitions by the same $\tol$ and specifying a consistent hyperparameter setting for digitization, i.e., $k=358$ and $\alpha=0.05$, respectively. The number of threads scheduled for JABBA is set the same as the partition number.  For each method, we report the number of symbols, reconstruction MSE, total runtime, and speedup. For JABBA, the number of partitions is set equal to the number of worker threads.

\begin{figure*}[ht]
\centering
\subfigure{\includegraphics[width=0.7\textwidth]{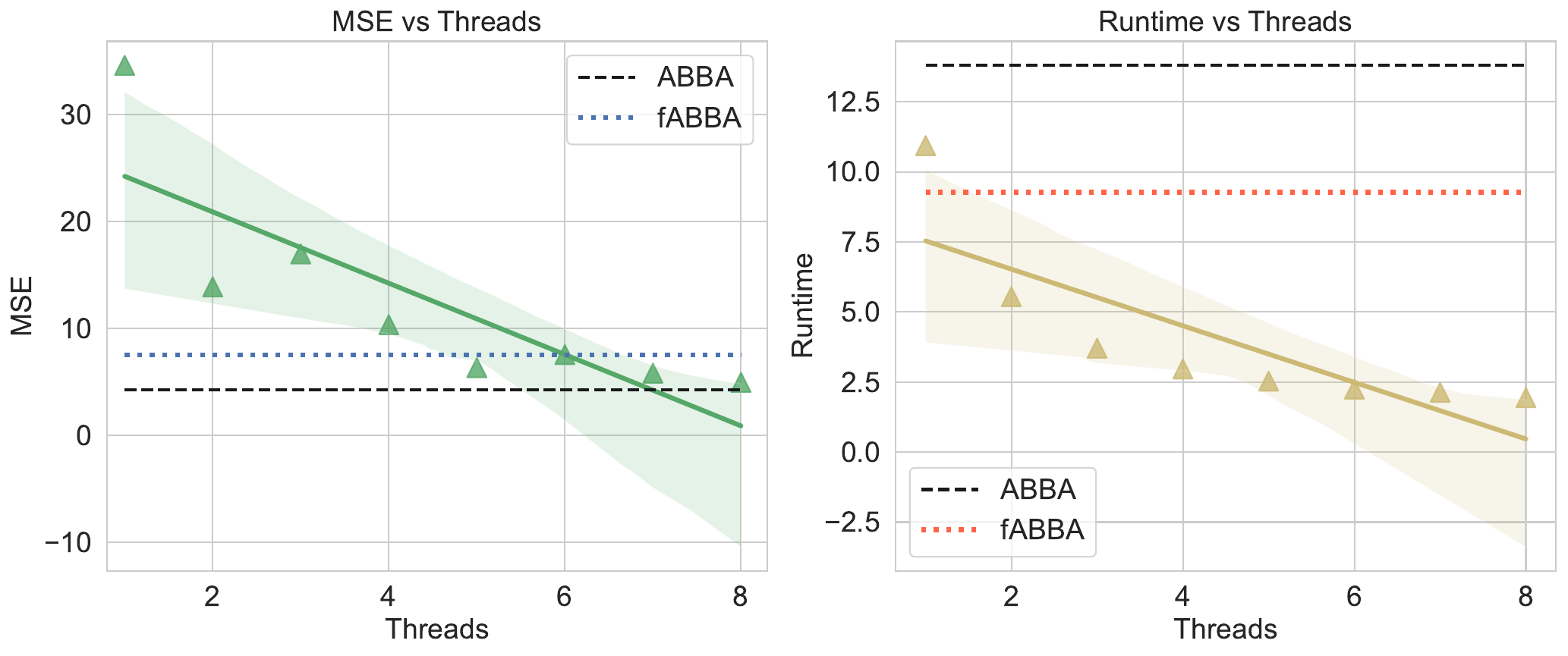}}
\caption{MSE and runtime of JABBA with varying number of partitions (the black line marks the result of ABBA and dashed line markes fABBA).}
\label{fig:gaussian_mse}
\end{figure*}

\begin{figure*}[ht]
\centering
\includegraphics[width=0.7\textwidth]{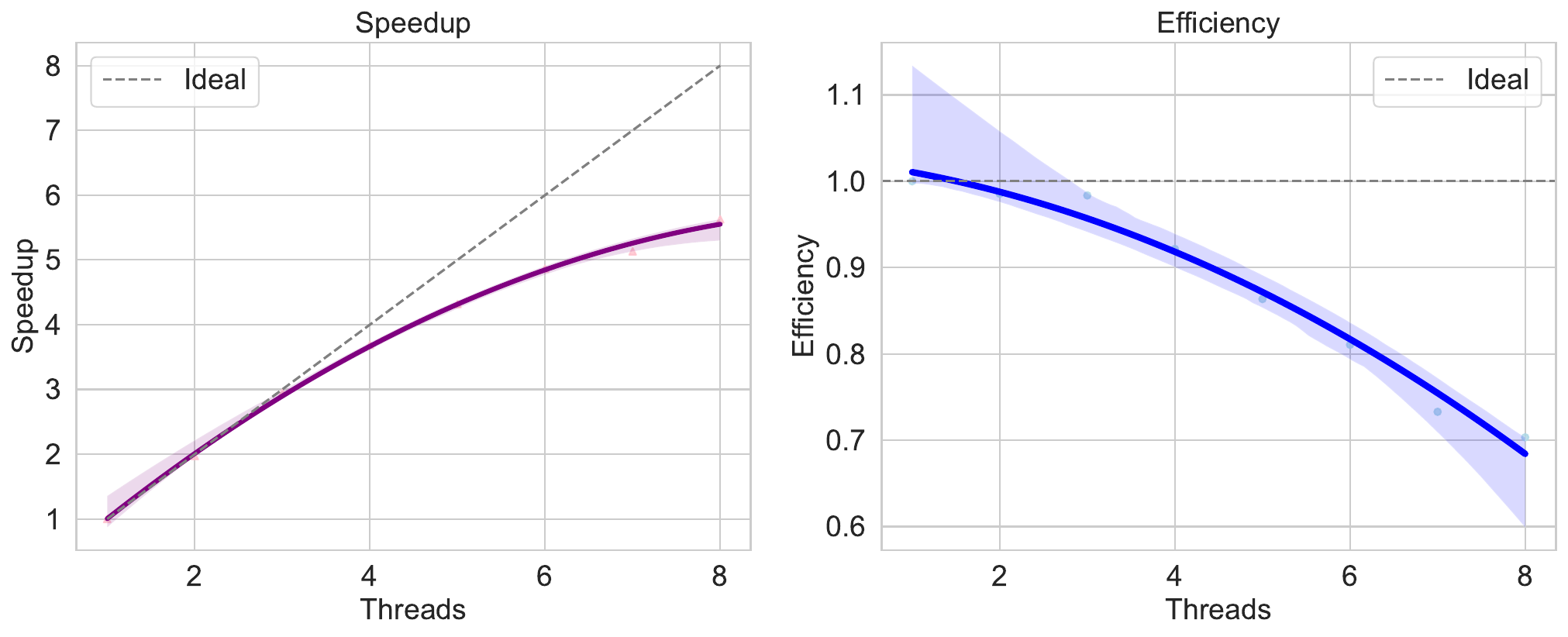}
\caption{Speedup and efficiency.}\label{fig:speedup}
\end{figure*}

The experimental result is exhibited in \figurename~\ref{fig:gaussian_mse} and \figurename~\ref{fig:speedup}, and we use serial ABBA and fABBA as baselines. In terms of runtime, JABBA achieves substantial speedup as the number of partitions increases. For example, using 8 threads reduces the runtime from 10.9 seconds to 1.94 seconds. Compared to the baselines, JABBA significantly outperforms both ABBA (13.8 seconds) and fABBA (9.27 seconds), demonstrating the effectiveness of the proposed parallelism in the compression stage. 
We can see a clear negative correlation between reconstruction error and the number of partitions. This can be explained by the increasing number of partition points used for reconstruction.  \figurename~\ref{fig:gaussian_mse} also shows that JABBA achieves similar performance against ABBA and fABBA regarding MSE while performing speedup by orders of magnitude.

The parallel speedup wand parallel efficiency ith $M$ workers are defined as
\begin{equation}
     \Phi(M) = \frac{\upsilon(1)}{\upsilon(M)}, \qquad E(M) = \frac{\Phi(M)}{M}.
\end{equation}
where $\upsilon(M)$ denotes the runtime using $M$ workers. 

Efficiency is important because it indicates whether the additional workers are effectively utilized. Without loss of generality, we only evaluate the speedup of Parallelism for JABBA as shown in \figurename~\ref{fig:speedup}. Although the current implementation keeps the digitization stage sequential, the overall runtime still scales well with the number of partitions. This indicates that the compression stage dominates the computational cost for large-scale time series and that domain decomposition effectively exposes the main source of parallelism. The speedup is close to linear for a small number of workers and remains above $5.6\times$ with eight workers, with a parallel efficiency of about $70\%$.  We can see the speedup $\Phi(M)$ scale almost linearly with the number of threads $M$. Since our algorithm is partially parallel in compression, which is hindered by the sequential part of the algorithm, that is, the digitization.  This phenomenon can be naturally explained by Amdahl's law which gives the theoretical speedup at a fixed workload where there are limits on the benefits one can derive from parallelizing a computation.

\section{Summary and future work}\label{sec:final}

We present a joint symbolic approximation method called JABBA, which enables an efficient parallel computing for large-scale time series symbolization. Unlike existing ABBA-based methods, which symbolize each time series independently, JABBA constructs a shared dictionary, thereby guaranteeing consistent symbols across series. The framework separates local compression from global digitization: compression is performed independently across multiple domains and can run in parallel, while digitization learns a common codebook from all compressed pieces. 

Our experiments on real time series datasets and synthetic large-scale time series show that JABBA substantially reduces runtime while retaining competitive reconstruction quality. The results suggest that symbolic approximation can be implemented as an extensible high-level data analytics primitive for temporal data. 

In future work, we plan to further improve the scalability of the proposed framework by extending it to distributed environments beyond shared-memory systems. In addition, a theoretical analysis of the approximation error and convergence properties of the proposed two-stage aggregation scheme remains an interesting direction.

\bibliographystyle{ACM-Reference-Format}
\bibliography{sample}

\end{document}